\documentclass[multphys,vecphys]{svmult}

\usepackage{makeidx}         
\usepackage{graphicx}        
\usepackage{multicol}        
\usepackage[bottom]{footmisc}

\makeindex             

\begin{document}

\title*{Nuclear Star Clusters in Edge-on Galaxies}
\author{Anil C. Seth\inst{1}\and
Julianne J. Dalcanton\inst{1,2}\and Paul W. Hodge\inst{1}\and Victor P. Debattista\inst{1,3}}
\authorrunning{Seth et al.}
\institute{University of Washington,
\texttt{seth@astro.washington.edu}
\and Alfred P. Sloan Research Fellow
\and Brooks Prize Fellow
}
%
%
\maketitle

From observations of edge-on, late-type galaxies, we present
morphological evidence that some nuclear star clusters have
experienced {\it in situ} star formation.  We find three nuclear
clusters that, viewed from the edge-on perspective, have both a
compact disk-like component and a spheroidal component.  In each
cluster, the disk components are closely aligned with the major axis
of the host galaxy and have bluer colors than the spheroidal
components.  We spectroscopically verify that one of the observed
multiple component clusters has multiple generations of stars.  These
observations lead us to suggest a formation mechanism for
nuclear star clusters, in which stars episodically form in compact
nuclear disks, and then lose angular momentum, eventually forming an
older spheroid.  The full results of this study can be found in
a forthcoming paper.

\section{Background}
\label{sec:1}

Nuclear star clusters are a common feature of dwarf elliptical and
spiral galaxies.  Surveys of both face-on bulgeless spirals (type Scd
and later) and dwarf ellipticals find that roughly 75\% have a single
bright star cluster as their nuclei\cite{boker02,cote06}.
The sizes of these nuclear clusters are similar to Galactic
globular clusters ($r_{eff} \sim 3$~pc), but they are
significantly brighter, with absolute $I$-band magnitudes of -8 to -16
\cite{boker02,boker04a}.  This luminosity is due both to their high
masses (typically a few\,$\times$10$^6$~M$_\odot$) and to the
presence of younger stellar populations \cite{walcher05,walcher06}. 

Nuclear clusters are interesting objects in a galaxy evolution
context.  A number of groups have recently shown that the masses of
nuclear clusters correlate with their host galaxy masses along the
same relation found for supermassive black holes, and thus appear to
be directly connected to the process of galaxy formation
\cite{ferrarese06,rossa06,wehner06}.  Furthermore, observations of
nuclear clusters may provide clues to the formation of unseen
supermassive black holes.  Lastly, nuclear clusters are possible
progenitors to massive globular clusters such as $\omega$\,Cen and
ultracompact dwarfs \cite{gnedin02,bekki04}.

Two scenarios have been suggested to explain the formation of nuclear
star clusters: (1) nuclear clusters form from multiple globular
clusters accreted via dynamical friction \cite{tremaine75}, and (2)
nuclear clusters form {\it in situ} from gas channeled into the center
of galaxies \cite{milosavljevic04}.  In this proceeding we present
evidence for the latter scenario.

\section{Results}

Our sample of galaxies consists of 14 nearby (2-20 Mpc), late-type
(Sbc+), edge-on spiral galaxies observed with HST/ACS as part of a
Cycle 12 snapshot program (sample details can be found in
\cite{seth05a}).  In these 14 galaxies, we identified 9 nuclear cluster
candidates.  

\begin{figure}[t]
\centering
\includegraphics[height=4.5cm]{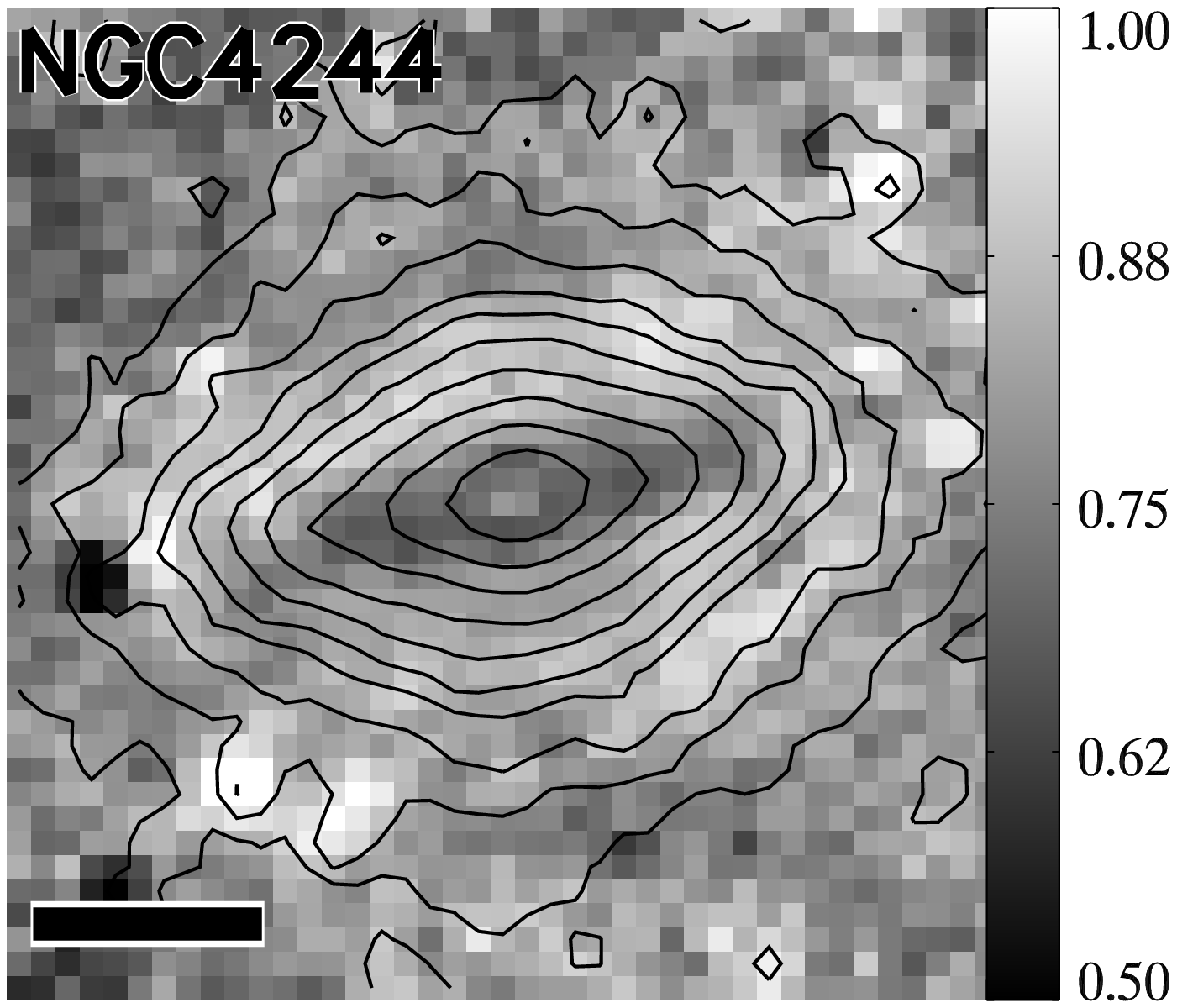}
\includegraphics[height=4.5cm]{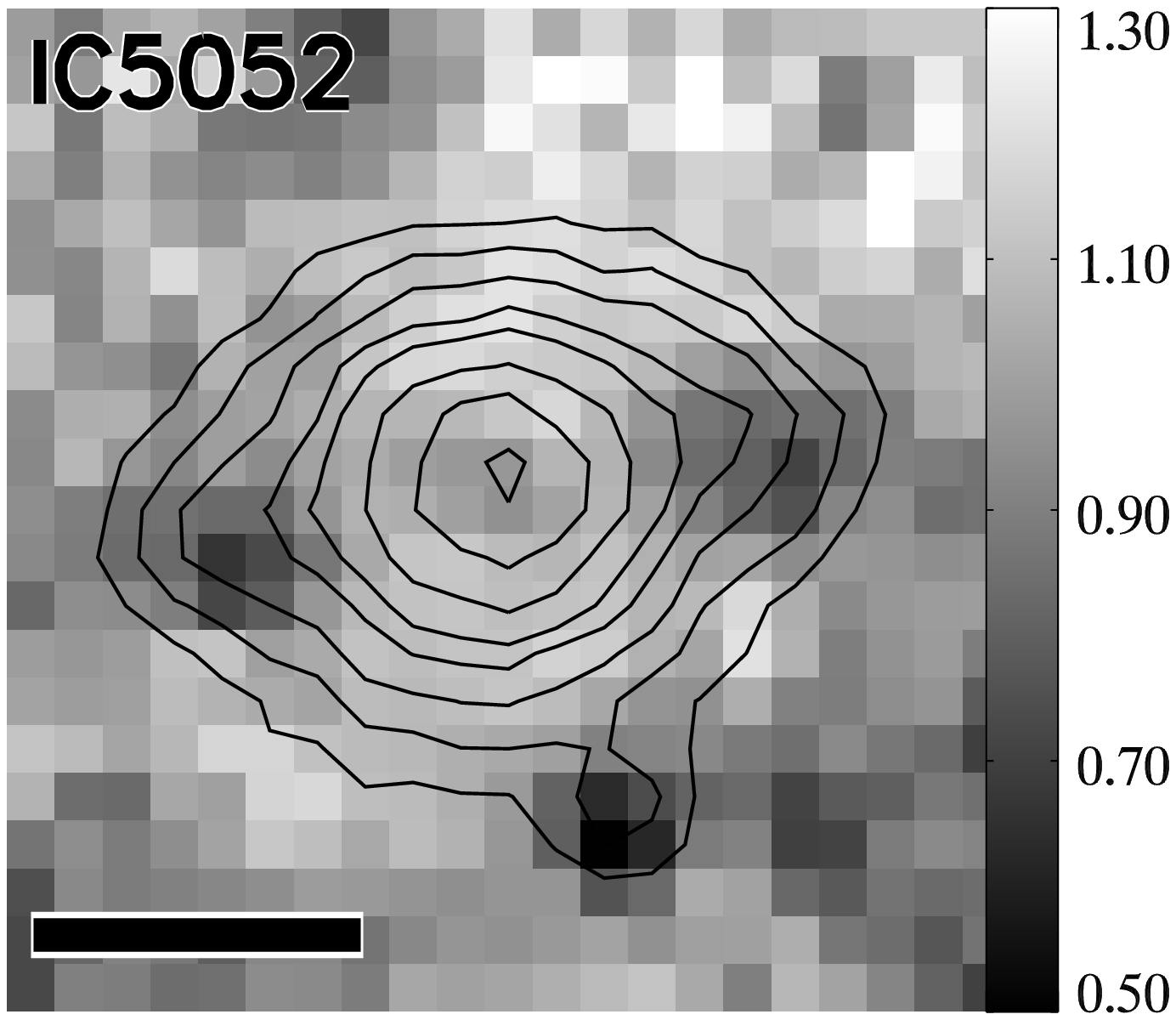}
\includegraphics[height=4.5cm]{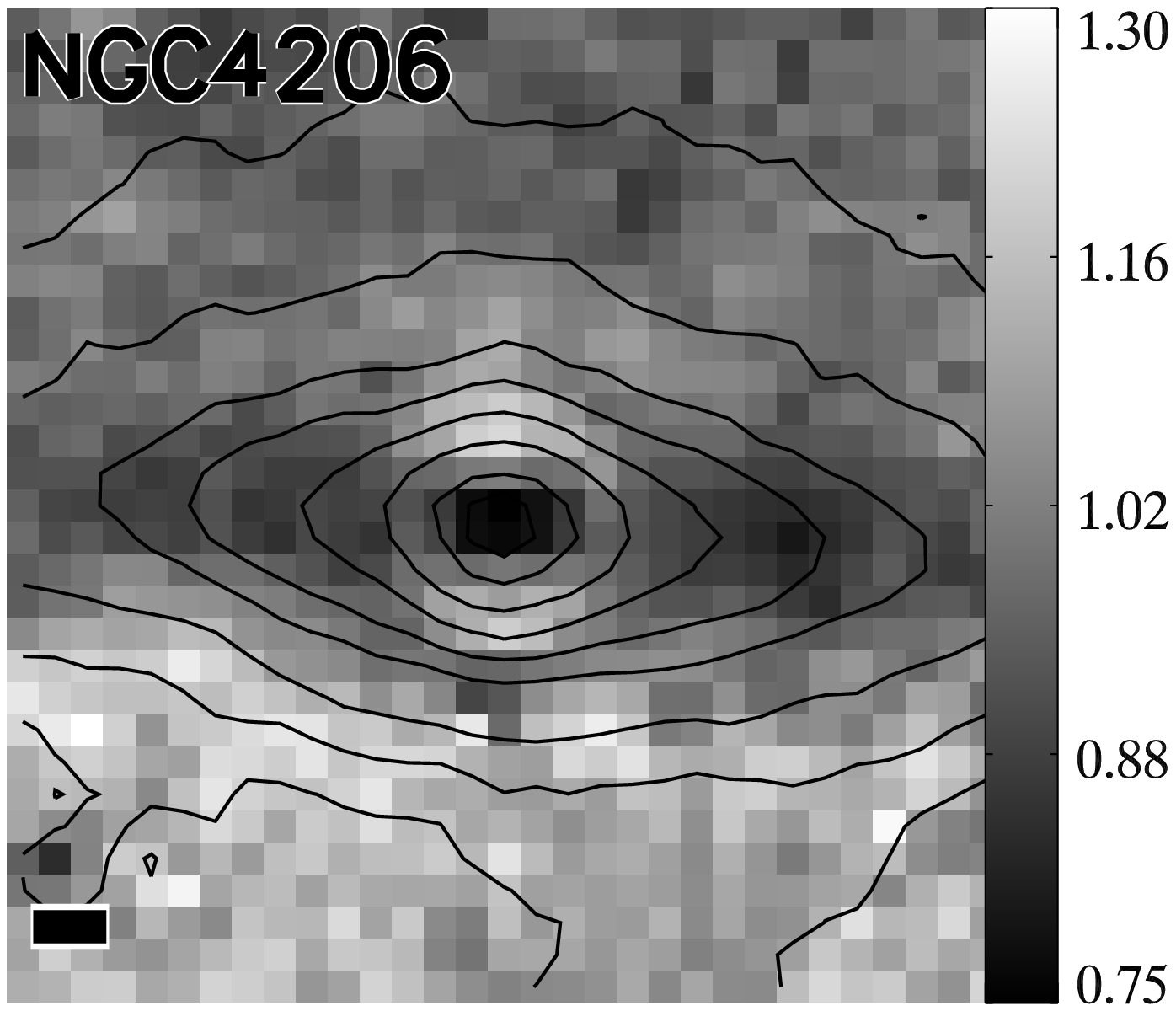}
%
%
\caption{Color maps of the three multi-component nuclear clusters
overlaid with contours showing the F606W brightness.  The colorbars
indicate the F606W-F814W in each cluster; dark colors indicate blue
regions.  Each image has been rotated so that the x-axis is parallel
to the major axis of the galaxy disk.  The black bar in the bottom
left corner indicates a length of 10 pc.}
\label{fig1}       
\end{figure}

\paragraph{Nuclear Cluster Morphologies and Luminosities} 

All nine of the detected cluster candidates were at least partially
resolved in the HST/ACS images.  Fits of convolved King profiles gave
effective (half-light) radii ranging from 1 to 20 pc, with most of the
clusters having effective radii between 1 and 4 pc.  This size range
is similar to what has been found previously for nuclear star clusters
in face-on, late-type galaxies \cite{boker04a} .  Furthermore, the
absolute $I$-band magnitudes are also similar to previously observed
nuclear clusters, ranging from -8 to -15 \cite{boker02}.

Three of the cluster candidates (in IC\,5052, NGC\,4206, and NGC\,4244)
have unusual morphologies.  As shown by the contours in
Figure~\ref{fig1}, these three candidates are elongated and appear to
have both a disk-like and spheroidal component, much like miniature S0
galaxies.  Fitting these clusters with both an exponential disk
component and an elliptical King profile component gave much smaller
residuals than single component fits.

The elongations and disk components of the three multi-component
clusters are aligned to within 10$^\circ$ of the major axis of the
edge-on galaxy disks (see Fig.~\ref{fig1}).  Previous studies of
nuclear clusters have focused on face-on galaxies, making detection of
similar multi-component clusters difficult.

\paragraph{Nuclear Cluster Stellar Populations} 

The color maps in Figure~\ref{fig1} show that the multiple
morphological components have clearly different F606W-F814W colors.
In each cluster, the disk components are bluer than the spheroid, with
a color difference $>$0.3 magnitudes.  This color difference is most
simply interpreted as a difference in age, with the disk being made of
younger stars than the spheroid.  Although the reddening is unknown,
based on Padova single-stellar population models in the ACS filters
\cite{girardi06}, the observed color difference implies that the
stellar ages of the disk are younger than $\sim$1~Gyr.

\begin{figure}[t]
\centering
\includegraphics[width=11cm]{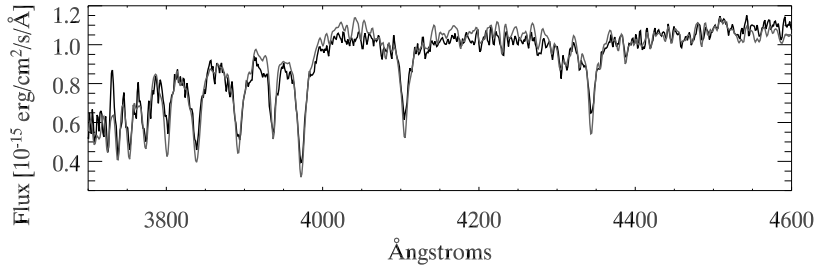}
%
%
\caption{Spectrum of the NGC\,4244 nuclear cluster obtained with the
Apache Point Observatory 3.5m telescope using the DIS spectograph
(black line).  The gray line shows the best-fitting two-age spectrum,
with stellar populations of 0.1 and 1 Gyr and a total mass of
3.5$\times$10$^6$~M$_\odot$.}
\label{fig2}       
\end{figure}

For the nuclear cluster in NGC\,4244, the nearest in our sample
(${\rm D} =4.4$~Mpc), we obtained a long-slit spectrum of the cluster using the
DIS spectrograph on the Apache Point Observatory 3.5m telescope
(Fig.~\ref{fig2}).  This spectrum verifies that multiple stellar
populations are present in the cluster, with the youngest component
having an age of $\sim$0.1~Gyr.  We fit the spectrum using
combinations of Bruzual \& Charlot models \cite{bruzual03} assuming
Z=0.008.

As would be expected from the color maps, the spectrum is much better
fit by multiple stellar populations than any single stellar
population.  In Figure~\ref{fig2} we show the best fitting two-age fit
with ages of 0.1 and 1~Gyr, and a mass of 3.3$\times$10$^6$~M$_\odot$, 5\%
of which is in the younger 0.1~Gyr component.  The luminosity of this
young component matches the disk luminosity of the best morphological
fit.  However, many different combinations of masses and ages fit the
data well, including a three-age model with a significant old (10~Gyr)
component, and a constant star formation rate model.  

\section{Summary and Discussion}

Three of the nine nuclear cluster candidates in our sample have young
disk components ($<$1~Gyr) in addition to an older spheroidal
component.  These disks cannot be formed by accretion of globular
clusters and must instead be formed {\it in situ} from gas accreted
into the nuclear regions.  The multiple stellar populations observed
in many nuclear clusters \cite{walcher06}, and the direct detection of
a molecular gas disk coincident with the nuclear cluster in IC~342
\cite{schinnerer03} provide additional evidence that nuclear cluster
formation is an ongoing process.  We propose a model for nuclear
cluster formation in which the stars in the cluster form episodically
in nuclear disks.  Such episodic star formation would naturally result from
stochastic accretion events and/or feedback from star formation.
Then, over time, the stars in the disk lose angular momentum and end up in a
more spheroidal component.  We are currently investigating the
mechanism by which the stellar disks could lose angular momentum.

%
%
%
%
%
%

%
%



\printindex

%
%
%

\end{document}